# A time-varying study of Chinese investor sentiment, stock market liquidity and volatility: Based on deep learning BERT model and TVP-VAR model[1]

ZHANG Chenrui[2], WU Xinyi, DENG Hailu, ZHANG Huiwei

*School of Economics and Finance, South China University of Technology*
*Guangzhou 510006, China*

*[Abstract] Based on the commentary data of the Shenzhen Stock Index bar on the EastMoney website from January 1, 2018 to December 31, 2019. This paper extracts the embedded investor sentiment by using a deep learning BERT model and investigates the time-varying linkage between investment sentiment, stock market liquidity and volatility using a TVP-VAR model. The results show that the impact of investor sentiment on stock market liquidity and volatility is stronger. Although the inverse effect is relatively small, it is more pronounced with the state of the stock market. In all cases, the response is more pronounced in the short term than in the medium to long term, and the impact is asymmetric, with shocks stronger when the market is in a downward spiral.*

*[Keywords] BERT model; TVP-VAR model; SZSE Component Index; investor sentiment; stock market liquidity; stock market volatility*

---

[1] This article is funded by the National Student Innovation and Entrepreneurship Training Program (Project No. 202110561076)

[2] Corresponding author, E-mail:charles.cheung@foxmail.com

# 1 Introduction

The relative stability of the stock market is the basis for its development and growth, which facilitates the entry of domestic and foreign investors and helps avoid the outbreak of financial crises. Abnormal stock market volatility will cause most investors to lose confidence in the market and also distort the stock price signal as an asset allocation indicator, which in turn will increase the systemic risk of the entire financial system. Compared to the mature stock markets in Europe and the U.S., the Chinese stock market is more volatile, and the 2014-2015 SSE Composite Index crash triggered a rethinking of the stock market and investors, the causes of which can be explained from two perspectives: behavioral finance and fractal market theory, the former highlighting the important influence of investors' behavior and sentiment, and the latter emphasizing the market liquidity factor. It is thus clear that the dynamics of investor sentiment and market liquidity have become important influences on market volatility. As stock trading has become more Internet-based and popular, the importance of exploring the interaction mechanism between investor sentiment, stock market liquidity and volatility has become increasingly important.

# 2 Review of the literature

In recent years, as stock market trading has become more internet-based and popular, the importance of exploring the mechanisms of investor sentiment on the market has become increasingly important in the field of behavioral finance research. By reading the literature review, this paper divides the research work into two parts: "investor sentiment measurement" and "relationship between investor sentiment and stock market", and further refines the direction of literature reading and research study based on this.

## 2.1 Investor sentiment

The direct survey method requires a lot of time and effort from the researcher and has a large bias with limited sample size and has been less used in recent studies. The market proxy method is often used to construct proxies for investor sentiment by selecting proxy variables such as closed-end fund discount and turnover rate and using principal component analysis. This method is still widely used because it is easy to implement, and the research effect is more significant. However, it also has the drawback that it can only indirectly reflect investor sentiment. The dictionary method, on the other hand, obtains a weighted investor sentiment indicator by comparing sentiment words appearing in financial texts with a dictionary. This method can directly reflect investor sentiment, which greatly facilitates the research on financial text analysis at home and abroad, but it also has the problem that sentiment indicators are greatly affected by the quality of dictionaries, and the workload of constructing dictionaries is tedious. With the improvement of computer computing power and the continuous update of algorithms in the field of natural language processing, many scholars began to borrow support vector machine (SVM), recurrent neural network (RNN) and long short-term memory unit (LSTM) model algorithms and use machine learning methods to directly extract investor sentiment from financial texts and construct indicators. Such research continues to improve accuracy with the improvement of machine learning modelling algorithms. Huicheng Liu uses bidirectional LSTM to encode news texts and search for contextual sentiment signal features; Yao Qin uses nonlinear autoregressive models to construct a two-stage recurrent neural network time-series-based prediction method to study investor sentiment. Currently, machine learning method is the most cutting-edge direction in the field of investor sentiment measurement, so this paper considers a machine learning approach to study investment sentiment.

Sentiment classification is a subtask in natural language processing tasks. It is essentially the process of analyzing, processing, generalizing and reasoning about subjective texts with emotional overtones. While this paper studies the analysis and processing of stock bar sentiment, sentiment classification is beneficial to the reasonable detection of stock market opinion monitoring.

With the development of computer hardware, a deeper and more complex deep learning model with a deeper network structure, BERT, was thus devised. Studies have found that BERT is more often used in media, political commentary research> For example, Guoshuai Zhang et al. (2020) used a BERT language model to analyse news published by the New York Times to predict future policy changes in the US in the short term; Jiaqi Hou et al. (2020) combined a BERT model with a two-way encoder of transformer to propose a new BERT-att and demonstrate that this model is more effective than the leading baseline model for use in areas such as security risk assessment. However, in the field of financial research, there are still few scholars using BERT model in their research, and fewer scholars have open-sourced their debugged and ready-to-use machine models for sentiment judgement. This research hopes to fill in the gaps and lacunae in this area so that future research can continue to advance and optimise in this direction. The research in this paper fine-tunes the BERT pre-training model, one of the most cutting-edge deep learning models in the current computer text processing field, to achieve a more suitable and reliable model for judging Chinese investors' text sentiment BERT model.

## 2.2  Market liquidity

Stock market liquidity refers to the phenomenon that the stock market executes a large number of transactions quickly and cheaply without causing significant changes in prices. Liquidity includes several aspects. One is the timeliness of transactions. Second, the low cost of the transaction. Third, the number of tradable stocks is huge. Fourth, the price is low. At present, there are many liquidity indicators used in the research, and different liquidity indicators can be constructed according to the needs of the research direction. Yang Chaojun (2008) proposed a new liquidity index based on the ideas of Amivest liquidity ratio and Hui heubel liquidity ratio. Zeng Zhijian and Luo Changqing (2008) studied the linkage relationship between the liquidity of the two markets by taking the turnover rate as the liquidity variable and using the data of Shanghai stock exchange for one year. It is found that there is no leading lag relationship between the monthly liquidity of stock market and bond market. Wang Yintian et al. (2010) studied the liquidity spillover effect of stock market and bond market, and constructed the liquidity index of liquidity measurement method based on price influence. Han Jinxiao et al. (2017) used crowin and Schultz (2012) price spread indicators to effectively measure the liquidity of China's stock market. In the stock market, turnover rate is the simplest liquidity index.

## 2.3 Market Volatility

Stock market volatility refers to changes in stock prices resulting from changes in expectations of the market due to changes in economic factors, policy factors and market participants' own factors. Stock market volatility is influenced by a variety of factors such as fundamentals, capital, and investor sentiment. For financial markets, volatility is one of the very important characteristics. It reflects changes in the returns of financial assets with each other. The more common measures of volatility in the literature include the variance or standard deviation of returns, the conditional variance of returns, the magnitude of shocks, etc. Engle (1982) distinguished between variance and conditional variance in his study and established the ARCH model, which opened up a new way of thinking in volatility research. GARCH model. Huang, Hainan and Zhong, Wei (2007) proposed the use of GARCH-like models to predict the return volatility of the SZCZ. Wen Dai (2018) uses the ARMA-GARCH models to empirically analyse the returns of specially treated classes of stocks. The existing literature on the study of stock market volatility reveals the aggregation, persistence and asymmetry of security price volatility and validates the applicability of ARCH models in modelling security price volatility. A variety of models currently proposed by domestic and international scholars consider stock market volatility from different aspects and can be selected based on the research needs, based on the actual situation. This paper refers to the realised volatilities in the CSMAR database to construct stock market volatility indicators.

## 2.4 Investor sentiment correlates with stock liquidity, stock volatility

In terms of the impact of investor sentiment on liquidity, Chen et al. (2009) studied the link between investor sentiment, market liquidity and investor trading behaviour and found that when investor sentiment tends to be pessimistic, more investors will choose to sell, reducing net market buying and hence market liquidity. Also individual investors are more responsive to changes in sentiment than institutional investors. Using Granger causality, Li Chunhong and Peng Guangyu (2011) find that investor sentiment is the Granger cause of stock liquidity, and that investor sentiment indirectly affects the capitalisation rate of the stock market and the level of economic growth by influencing stock liquidity. Wang Danfeng and Liang Dan (2012) find that changes in investor sentiment increase market liquidity and the effect on the liquidity premium of expected stock returns becomes significant. Liu et al. (2016) find that investor sentiment positively affects market liquidity and that stock market liquidity is relatively weaker the higher the investor's cognitive ability regarding stock market information. In addition, the study found that there is a difference in the degree of influence of positive and negative investor sentiment on stock market liquidity. Yang et al. (2016) studied the impact of local preference and investor sentiment on the stock market and found that the positive impact of local concern on stock trading volume was greater under positive sentiment than under negative sentiment. Using the TVP-SV-SVAR model to analyse the impact of investor sentiment, market liquidity on stock market bubbles, Shi Guangping et al. (2016) found that the impact of investor sentiment on stock market liquidity and the impact of optimism in a bull market is more significant than the impact of pessimism in a bear market. The above literature has examined the impact of investor sentiment on stock market liquidity from different perspectives, and to some extent has identified the differential impact of different types of sentiment on the stock market and its differential performance under different periods.

In terms of the impact of investor sentiment on stock market volatility, Lee et al. (2002) found that investor sentiment has a significant impact on stock market volatility, where positive sentiment diminishes market volatility and negative sentiment increases market volatility. Zhang Zongxin and Wang Hailiang (2013) used principal component analysis to construct an investor sentiment index to analyse the relationship between investor sentiment and stock market volatility and found that investor sentiment has a significant positive impact on market returns and volatility, with higher levels of sentiment leading to greater changes in stock market returns, greater deviation of stock prices from intrinsic value and greater stock market volatility. Hu, Changsheng and Chi, Yangchun (2013) find that both rational and irrational investors are the creators of market volatility and investor sentiment can cause the market to deviate from the rational framework and lead to abnormal volatility. Lu Jianqing and Chen Mingzhu (2013) proposed the hypothesis of "expected emotional spillover effect", suggesting that the psychological expectations of investors at the micro level stimulate the herding effect, which induces the "stock market resonance effect" at the macro level and exacerbates stock market volatility. Ba Shu Song (2016) points out that financing and financing will fuel investor sentiment and have an exacerbating effect on market volatility. From different perspectives, the above literature confirms that investor sentiment is one of the systematic risk factors in the stock market and an important factor affecting the volatility of stock market returns.

# 3 Data Processing & Modelling

## 3.1 Selection of data

Considering the selection of the research object, it is necessary to have a better overview of the whole A-share market, which can reflect the actual situation of investor sentiment in an all-round way. In this paper, the stock bar of EastMoney[3] was selected as the source of textual data, and realized volatility ($RV_t$) and turnover rate ($Turn_t$) were chosen as variables to study the impact of investor sentiment.

### 3.1.1 Selection of text data

This article crawls the stock bar data from January 1, 2018 to December 31, 2019 for the EastMoney Website stock bar. A number of individual stocks as well as index stock bar data were compared horizontally, and finally the stock bar of SZCZ was selected. Compared to the SSE Composite Index and other index stock bar data, the SZCZ stock bar has the characteristics of suitable data volume, large number of viewers and strong representation. Secondly, the SZCZ contains data on 500 high-quality companies listed on the SZSE, covering various industries. Therefore, SZCZ can reflect the situation of the stock market very well. In terms of text processing, the text analysis standard is followed and non-text items such as duplicate data, tables and HTML tags in the data are removed.

---

[3] According to a 2018 analysis by iResearch, EastMoney is the top financial website in China. It has 78 million hours of active browsing time per month and accounts for 45% of the market share, which is higher than the remaining nine companies in the top 10 companies combined.

**Table 1 Selected constituent stocks of the SZCZ as at 1 January 2018**

| Ranking | Code | Abbreviations | Weighting (%) | Total equity (trillion) |
|---|---|---|---|---|
| 1 | 000333.SZ | Midea Group | 3.63 | 65.61 |
| 2 | 000651.SZ | Gree Electric | 3.10 | 60.16 |
| 3 | 000858.SZ | Wuliangye | 2.20 | 37.96 |
| 4 | 000725.SZ | BOE A | 2.19 | 347.98 |
| 5 | 000002.SZ | Vanke A | 1.88 | 110.39 |

Note: Data sourced from the Flush iFind database.

**Table 2 Summary information on daily comment statistics**

| Date | Comments | Readership | Number |
|---|---|---|---|
| 01/02/2019 | There is no doubt that today is another opportunity to dump stocks that have risen tremendously. Must face ups and downs with a calm mind. | 2115 | 0 |
| 01/02/2019 | Northbound foreign capital flowed into A-shares with a net inflow of over RMB 310 billion. That is good news. | 1159 | 2 |
| 01/02/2019 | A fundamental change in the stock market's internal and external environment. Don't look at the stock market through the eyes of a stale bear market. The stock market is opening up in a big way share prices must gradually rise. | 921 | 0 |

As the market capitalisation of stocks is constantly changing, the constituent stocks are also constantly changing. Table 1 shows some of the constituent stocks of the SZCZ as at 1 January 2018, but the constituent stocks did not change significantly during the study period of this paper. Therefore, the impact of changes in constituents can be disregarded. In addition, the information we crawled for the daily stock bars included the time, comment content, number of reads and comments, etc., as shown in Table 2.

The statistical information obtained from Table 3 shows that the average comment headline length per day is 62.7 characters and that the content is right-handedly distributed. In contrast to traditional (relatively short) comments, some of the more wordy stock comments in the stock bar data are often copied and pasted from other sources, such as news reports and analyst reports. Here we use a simple process to eliminate these potentially influential outliers, retaining only those comments with fewer than 150 characters. In addition, given the heterogeneity of daily comment counts and the effect of varying readership on the results, we chose to sort the data by readership, selecting the top 50 data items per day from highest to lowest for the study.

**Table 3 Basic statistical description of the pre-processed stock bar comment text data**

|  | Mean | S.D | Skewness | Min | Max | Count |
|---|---|---|---|---|---|---|
| Data length (characters) | 62.6754 | 169.0293 | 7.9438 | 2 | 2921 | 39784 |

Note: This table provides summary statistics for the length of comments per day. Sample means, standard deviations (S.D.), maximum and minimum numbers and total number of comments for the variables. The sample contains data on SZCZ stock bar comments for the sample period 1 January 2018 to 31 December 2019.

### 3.1.2 Selection of stock trading data

This paper selects the SZCZ as the research object, with the help of its daily trading market index, including the data of total market capitalisation, market capitalisation in circulation, total equity, share capital in circulation, turnover rate, P/E ratio and P/N ratio amount on that day. In this paper, the data is obtained through Tushare, an open source Python data API, and the results returned are of the Pandas.DataFrame data type.

**Table 4 Basic statistical description of the indicators for 2018-2019**

|  | Mean | S.D | Skewness | Min | Max |
|---|---|---|---|---|---|
| Exchange Rate (%) | 1.2565 | 0.5268 | 2.3546 | 0.66 | 3.99 |
| Volatility (%) | 1.1936 | 1.0730 | 3.6074 | 0.21 | 12.02 |

As can be seen from Table 4, the average value of stock market volatility and turnover rate is around 1.2% and both show a right-skewed distribution, where stock market volatility is high and the difference between the maximum and minimum values is about 10% which shows that the volatility is more violent.

## 3.2 Model construction

### 3.2.1 Deep learning model design based on financial corpus

Word Embedding is widely used for pre-training work in NLP (Natural Language Processing) in the era of deep learning. When training with a deep learning model, the trained subsets are transformed into word vectors as the input layer of the neural network. During the training of a deep learning model, the level of results depends heavily on the size of the training set, with larger training sets producing better word vectors. Currently, the vast majority of task models in the field of natural language processing use trained word vectors. During the training of word vectors, the word vectors ignore the contextual ideographs and when words have multiple meanings, they often correspond to the same word vectors. Therefore, in 2018 Devlin et al. (2019) proposed the pre-trained language model BERT, which topped the list of 11 NLP tasks and was a big step forward in the field of NLP. The structure of the BERT model is shown in Figure 1, where $E_1, E_2, \ldots, E_N$ are the input characters of the model, and the input characters are obtained through a bi-directional Transformer feature extractor to obtain text features, and the input characters are trained to output the corresponding vectors $T_1, T_2, \ldots, T_N$.

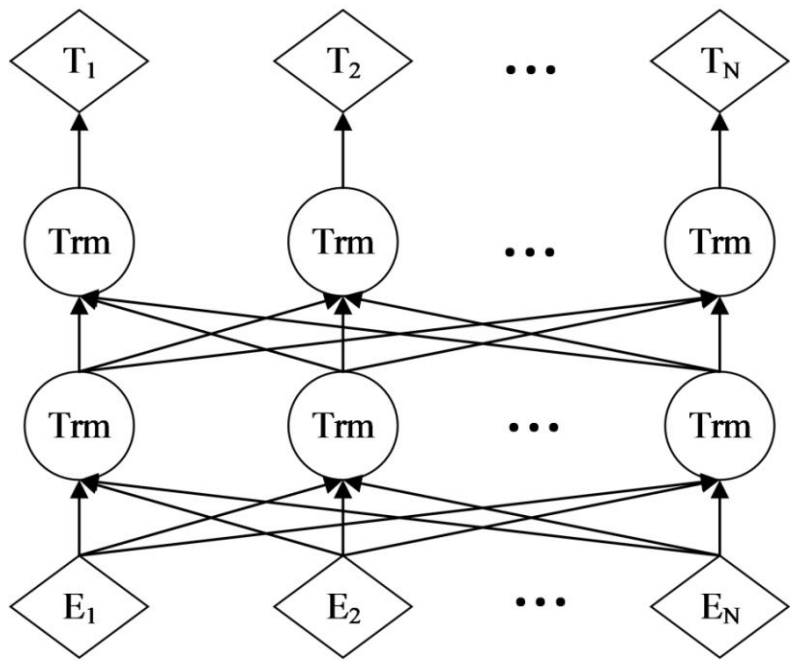

**Figure 1 BERT model structure**

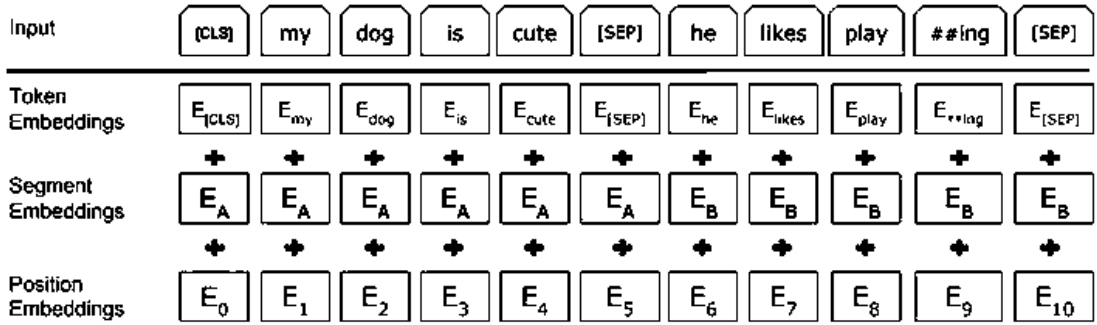

**Figure 2 How the BERT model is trained**

As shown in Figure 2, BERT is a model where all layers are trained with contextual semantics, and its input consists of three vectors of Token Embeddings, Segment Embeddings and Position Embeddings, while BERT uses a masked MLM, which masks a part of a word, similar to fill-in-the-blank, and then predicts the masked model, iterating through to achieve contextual training.

BERT is essentially a two-stage NLP model. The first stage is called pre-training, where an existing unlabelled corpus is used to train a language model. In order to improve the accuracy in prediction here, the Chinese BERT pre-training model based on full word coverage (wwm), Chinese-BERT-wwm, released by Joint Laboratory of HIT and iFLYTEK Research (HFL), is used as the pre-training model. after the pre-training, BERT can be used for the task of sentiment recognition of financial entities, which will be used as the output of the next network in sentiment analysis. Depending on the specificity of the financial text, a second stage of fine-tuning the financial text with sentiment annotation locally can be trained to produce a model with higher accuracy in the financial domain.

### 3.2.2 Model training

In this paper, we use the English_roberta_wwm_large_ext_L-24_H-1024_A-16 (24-layer, 1024-hidden, 16-heads) pre-trained model published by IIT Xunfei Joint Lab, which uses a 24-layer Transformer with a hidden layer dimension of 1024 and a total model size of 330MB. The model is trained with a batch size of 16, a learning rate of 2e-5, and a max seq length of 128. The emotions are labeled using three categories, with 0 indicating negative emotions, 1 indicating neutral emotions, and 2 indicating positive emotions. indicates positive sentiment.

Among the data crawled from January 1, 2018 to December 30, 2019, 4,000 data were randomly selected for the annotation of artificial emotions, and the annotated data were divided into a training set and a test set in the ratio of 8:2 for training. As shown in Table 5, the data processed included Sentiment 0 for negative, 1 for neutral, 2 for positive, Positive_Pro for the probability of the text belonging to positive sentiment, and Negative_Pro for the probability of belonging to negative sentiment.

Table 5 Selected sample data

| Date | Comments | Sentiment | Positive_Pro | Negtive_Pro |
|---|---|---|---|---|
| 12/21/2018 | A once-in-a-decade opportunity to double your wealth. Do you want to achieve wealth freedom? Take the plunge now and make your dreams come true. | 2 | 0.99148 | 0.00852046 |
| 12/21/2018 | Previous couplet: Borrow from the east wall to make up for the west wall and run out of money year after year. The next couplet: borrowing new to pay off old, losing money year after year. | 0 | 0.0499047 | 0.950095 |
| 12/21/2018 | How far is A-share from collapse? | 0 | 0.042031 | 0.957969 |

### 3.2.3 Investor sentiment indicators based on stock reviews

Drawing on the method proposed by Antweiler and Frank (2004) for constructing bullish indicators based on the classification of stock bar posts.

$$Sent_t = \frac{M_t^{pos} - M_t^{neg}}{M_t^{pos} + M_t^{neg}} \quad (1)$$

In equation (1) $M_t^c = \sum_{i \in D(t)} \omega_i \chi_i^c$ represents the sum of the weighted number of messages of type c∈{pos,neu,neg} over a period of time *D(t)* where *pos* represents positive sentiment, *neg* represents negative sentiment, *neu* represents neutral sentiment and $\chi_i^c$ is the indicator variable. In particular, when the weights are all equal to 1, $M_t^c$ is equal to the total number of messages of type $c$ over time $D(t)$. The stock comment sentiment indicator $Sent_t$, between -1 and 1, expresses the relative bullishness of investors and is independent of the total number of posts. Figure 3 shows that sentiment is predominantly negative, with negative sentiment being concentrated in 2018, which is associated with the one-sided downward trend in 2018.

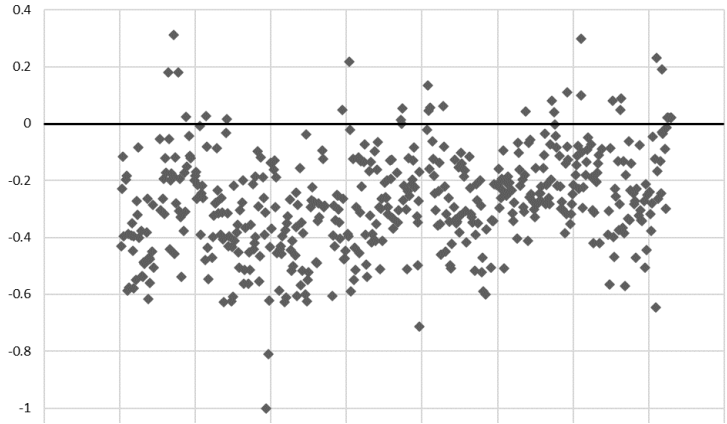

**Figure 3 Scatterplot of the distribution of emotions**

# 4 Empirical analysis

## 4.1 ADF

The three indicators, investor sentiment ($Sent_t$), realised volatility ($RV_t$) and turnover rate ($Turn_t$), are first tested for stationarity. The experimental results show that the series of investor sentiment ($Sent_t$) and realised volatility ($RV_t$) is integrated of order 0, while the series of turnover rate ($Turn_t$) is integrated of order 1 at the 1% significance level.

**Table 6 Summary of the results of the tests for the stability of variables**

|  | Inspection forms $(C,T,L)$ | t-statistic | p-value |
| --- | --- | --- | --- |
| $Sent_t$ | (1,0,1) | −10.89045 | 0.0000*** |
| $RV_t$ | (1,0,2) | −6.132106 | 0.0000*** |
| $dTurn_t$ | (1,0,4) | −14.31729 | 0.0000*** |

$(C,T,L))$ denote the intercept term, trend term and lag order respectively in the ADF test model. *** denotes significant at 1% confidence level

## 4.2 TVP-VAR modelling

The TVP-VAR model assumes that the variance of the coefficients and error terms is time-varying, satisfying a first-order stochastic wandering process. The time-varying nature allows the model to better capture possible, non-linear, asymptotic linkage changes or structural shifts in investor sentiment and stock market volatility. However, time-variability also makes it difficult for traditional likelihood function methods to reliably estimate parameters, and Nakajima (2011) proposes to overcome the shortcomings of traditional estimation methods by means of Markov Monte Carlo algorithms (MCMC). Firstly, the prior distribution of the time-varying parameters is set empirically. Secondly, a random sample of state vectors is drawn iteratively by the Markov Monte Carlo algorithm. When the relevant conditions are satisfied, the state transfer matrix of the Markov chain will converge to a stable probability distribution. The set of samples selected after convergence and satisfying the smooth distribution can effectively simulate the conditional posterior distribution of the time-varying parameters, and thus achieve statistical inference of the parameters.

The optimal lag order of the VAR model is judged to be 3rd order based on the results of the synthesis of the six judged indicator criteria (LogL, LR, FPE, AIC, SC, HQ). Drawing on Nakajima (2011) and Zheng Tingguo et al. (2018), this paper develops the following time-varying parametric vector autoregressive (TVP-VAR) model for investor sentiment ($Sent_t$), realized volatility ($RV_t$) and first-order difference series of turnover rates ($dTurn_t$).

$$A_t y_t = F_{1t} y_{t-1} + F_{2t} y_{t-2} + F_{3t} y_{t-3} + u_t, \quad u_t \sim N(0, \Sigma_t \Sigma_t) \qquad (2)$$

The vector $y_t$ is a $3 \times 1$ dimensional column vector ($Sent_t$, $RV_t$, $dTurn_t$). $F_{1t}$、$F_{2t}$、$F_{3t}$ is a $3 \times 3$ dimensional matrix of time-varying coefficients. The random perturbation term $\varepsilon_t$ is a $3 \times 1$ dimensional column vector reflecting the structural shocks. $\Sigma_t$ is a diagonal matrix with the standard deviation ($\sigma_{1t}, \sigma_{2t}, \sigma_{3t}$) of the random perturbation term as an element. $A_t$ is a lower triangular matrix with diagonal element 1 reflecting the simultaneous relationships between the different variables influenced by the structural shocks, which can be obtained by recursive identification.

Deforming equation (2) and adding the intercept term gives the final form of the equation.

$$y_t = C_t + B_{1t}y_{t-1} + B_{2t}y_{t-2} + B_{3t}y_{t-3} + e_t, \quad e_t \sim N(0, A_t^{-1}\Sigma_t\Sigma_t A_t'^{-1}) \quad (3)$$

Where $B_{it} = A_t^{-1}F_{it}$ ($i = 1,2,3$), stack the elements of $B_{it}$ in rows to form the $27 \times 1$ dimensional column vector $\beta_t$ and and stack the lower triangular elements on the non-diagonal of $A_t$ in rows to form the column vector $a_t$. $e_t = A_t^{-1}\Sigma\varepsilon_t$ ($\varepsilon_t \sim N(0, I_3)$), $h_t = (h_{1t}, h_{2t}, h_{3t})'$, and satisfy $h_{jt} = log\sigma_{jt}^2$ ($j = 1,2,3$).

Assume that all the parameters in equation (3) satisfy a first-order random walk process and that the error term follows the following distribution.

$$\beta_{t+1} = \beta_t + u_{\beta t}, \quad a_{t+1} = a_t + u_{at}, \quad h_{t+1} = h_t + u_{ht}$$

$$\begin{pmatrix} \varepsilon_t \\ u_{\beta t} \\ u_{at} \\ u_{ht} \end{pmatrix} \sim N\left(0, \begin{pmatrix} I & 0 & 0 & 0 \\ 0 & \Sigma_\beta & 0 & 0 \\ 0 & 0 & \Sigma_a & 0 \\ 0 & 0 & 0 & \Sigma_h \end{pmatrix}\right)$$

$$(\Sigma_\beta)_i^{-2} \sim Gamma(20, 10^{-4}), \quad (\Sigma_a)_i^{-2} \sim Gamma(4, 10^{-4}),$$

$$(\Sigma_h)_i^{-2} \sim Gamma(4, 10^{-4})$$

Referring to the model setup of Tingguo Zheng (2018), the following initial states are assumed.

$$\beta_0 \sim N(\mu_{\beta_0}, \Sigma_{\beta_0}), \quad a_0 \sim N(\mu_{a_0}, \Sigma_{a_0}), \quad h_0 \sim N(\mu_{h_0}, \Sigma_{h_0})$$

$$\mu_{\beta_0} = \mu_{a_0} = \mu_{h_0} = 0, \quad \Sigma_{\beta_0} = \Sigma_{a_0} = \Sigma_{h_0} = 10 \times I$$

## 4.3 Granger causality test

Granger causality tests are conducted on investor sentiment ($Sent_t$), realised volatility ($RV_t$) and first-order difference series of turnover rates ($dTurn_t$). The lag order is consistent with the TVP-VAR model (3rd order) and the experimental results are as follows.

**Table 7 Granger causality test table**

| Original assumptions | F-statistic | p-value |
|---|---|---|
| $Sent_t$ is not the Granger cause of the change in $dTurn_t$. | 2.48165 | 0.0603 |
| $dTurn_t$ is not the Granger cause of the change in $Sent_t$ | 3.14444 | 0.0250 |
| $RV_t$ is not the Granger cause of the change in $dTurn_t$ | 9.50156 | 4.E-06 |
| $dTurn_t$ is not the Granger cause of the change in $RV_t$ | 1.16143 | 0.3240 |
| $RV_t$ is not the Granger cause of the change in $Sent_t$ | 0.87328 | 0.4548 |
| $Sent_t$ is not the Granger cause of the change in $RV_t$ | 2.84445 | 0.0373 |

The test results indicate that the original hypothesis of "investor sentiment ($Sent_t$) is not a Granger cause of change in turnover rate ($dTurn_t$)" is rejected at the 10% significance level. At the 5% significance level, the original hypothesis that "the turnover rate ($dTurn_t$) is not a Granger cause of the change in investor sentiment ($Sent_t$)" and "investor sentiment ($Sent_t$) is not a Granger cause of the change in realized volatility ($RV_t$)" is rejected. The original hypothesis of "investor sentiment ($Sent_t$) is not the Granger cause of changes in realized volatility ($RV_t$)". At 1% significance level, the original hypothesis of "realized volatility ($RV_t$) is not a Granger cause of change in turnover rate ($dTurn_t$)" is rejected. That is, the pre-existing changes in investor sentiment can effectively explain the fluctuations in turnover and volatility. Pre-existing changes in turnover can effectively explain pre-existing changes in investor sentiment. The prior period change in volatility can effectively explain the volatility of turnover.

## 4.4 Analysis of experimental results

Set the number of iterations of the Markov Monte Carlo algorithm to 11,000. The first 1,000 "pre-burn" samples were discarded and the last 10,000 samples that satisfied the smooth distribution were used for estimation. The results of the estimation of the TVP-VAR model based on the Markov Monte Carlo algorithm are as follows.

Table 8 Table of estimation results of TVP-VAR model based on MCMC algorithm

| Parameters | Ave. | S.D | 95% confidence interval | Geweke diagnostic values | Invalid impact factor |
|---|---|---|---|---|---|
| $(\Sigma_\beta)_1$ | 0.0023 | 0.0003 | [0.0018,0.0029] | 0.864 | 28.76 |
| $(\Sigma_\beta)_2$ | 0.0023 | 0.0003 | [0.0018,0.0029] | 0.928 | 21.76 |
| $(\Sigma_a)_1$ | 0.0068 | 0.0021 | [0.0038,0.0113] | 0.900 | 99.99 |
| $(\Sigma_h)_1$ | 0.0064 | 0.0028 | [0.0033,0.0135] | 0.204 | 168.34 |
| $(\Sigma_h)_2$ | 0.7354 | 0.0841 | [0.5844,0.9123] | 0.374 | 45.01 |

The experimental results show that all estimated parameter values fall within their 95% confidence intervals and that the Geweke diagnostic values for all parameters do not exceed the critical value at the 5% significance level. In other words, the Markov chains of all parameter estimates are satisfying convergence. In summary, it can be seen that the parameter estimates possess validity characteristics.

### 4.4.1 Impulse response analysis with different lags

Referring to the studies of Yang, Wenqi et al. (2020) and Ren, Yongping et al. (2020), lags of 1 day, 7 days (one week) and 14 days (two weeks) are chosen as proxies to investigate the effects of unit shock variables on the formation of the shocked variables in the short, medium and long run, respectively.

- **Mechanisms of influence between investor sentiment and stock market liquidity at different lags**

The impulse response of investor sentiment to stock market liquidity indicators is constant and positive throughout the study time period, with some time-varying effects. The impact of shocks from changes in investor sentiment on stock market liquidity increases slowly with a lag of 1 period between June 2018 (around t=100) and July 2019 (around t=360). That is, when investors are confronted with changes in the equity market and after a period of confirmation of a bear market (around t=20), uncertainty about the future movement and duration of the market as well as risk averse preferences may lead them to trade frequently and thus make the market overreact. This effect continues until the investor enters a bull market (around t=246) after a period of confirmation.

In addition, the effect of lag 1 (short term) is significantly stronger than the effect of lags 7 and 14 (medium and long term). This also reflects the information efficiency of the stock market, in line with the findings of Yang, Wenqi (2020). Conversely, although the market liquidity indicator has a relatively small effect on investor sentiment, its time-varying effect is more significant (more strongly influenced by changes in the state of the stock market). The results of the experiment are shown in Figure 4.

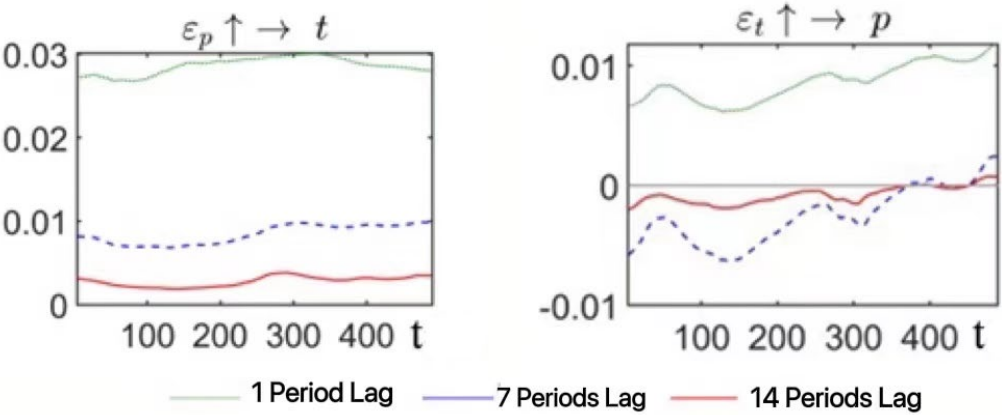

**Figure 4 Impulse response between investor sentiment and stock market liquidity at different lags**

- **Mechanisms affecting the relationship between investor sentiment and stock market volatility at different lags**

The results of the experiment are shown in Figure 5. The impulse response of investor sentiment to stock market volatility with a lag of 1 period is constant negative over the study time period and the shock is stronger in bear markets, with a weak effect in the medium to long term.

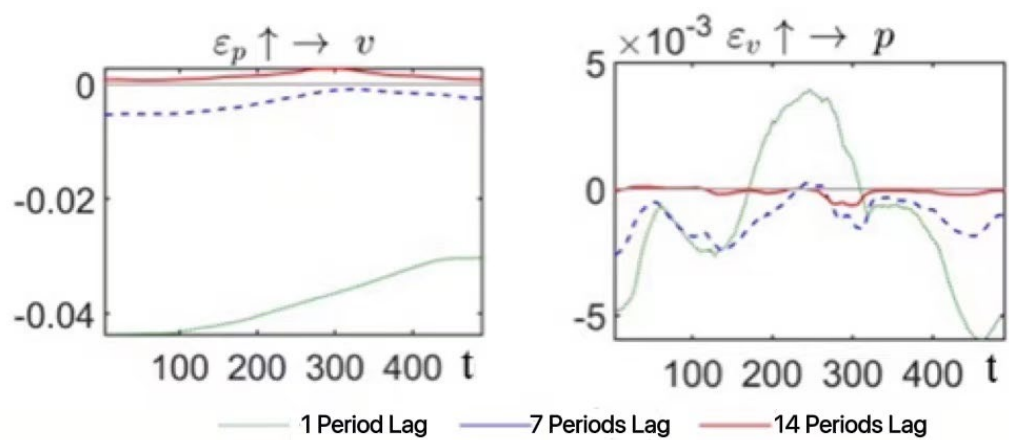

Figure 5 Impulse response plots between investor sentiment and stock market volatility at different lags

### 4.4.2 Impulse response analysis at different time points

The results of the experiment are shown in Figure 5. The impulse response of investor sentiment to stock market volatility with a lag of 1 period is constant negative over the study time period and the shock is stronger in bear markets, with a weak effect in the medium to long term.

- **The mechanism of influence between investor sentiment and stock market liquidity at different points in time**

The response of investor sentiment to shocks to stock market liquidity indicators is consistently positive at all three points in time, and the short-term impact is stronger than the long-term impact, with the impact of sentiment on stock market liquidity significantly weaker with a 3 period lag. The impact of investor sentiment on stock market liquidity is slightly stronger during bull markets. The experimental results support the notion that sentiment has an asymmetric impact on stock markets. Optimistic sentiment is more likely to drive stock prices upwards away from fundamental values, which in turn tends to lead to frequent trading and the creation of stock price bubbles. A possible transmission mechanism is that when investors observe high trading activity in the market and an initial increase in the turnover rate of stocks, their sentiment tends to rise and follow suit, i.e. a herding effect. The irrational decisions of individuals can easily evolve into convergent market sentiment and actions, resulting in frequent trading, deviations in asset prices and market volatility. However, when investors process information and make rational judgments, the impact of changes in stock market liquidity will be lessened. It is also worth noting that the impact of changes in market liquidity on investor sentiment is significantly stronger in bear markets than in bull markets. The results of the experiment are shown in Figure 6.

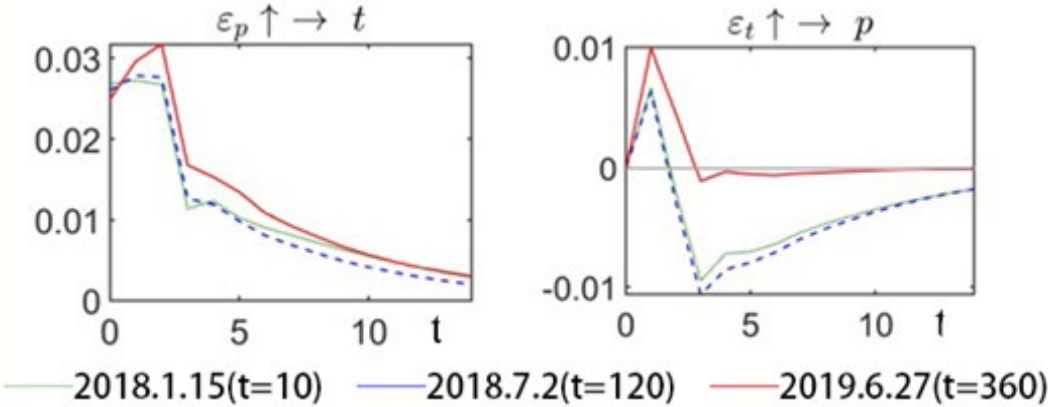

**Figure 6 Impulse response between investor sentiment and stock market liquidity at different points in time**

- **Mechanisms affecting the relationship between investor sentiment and stock market volatility at different points in time**

The results of the experiment suggest that investor sentiment has a stronger impact on stock market volatility indicators in bear markets than in bull markets, and that the magnitude of the impact gradually weakens after repeated changes in the latter five days. This may be related to the ability of investors to receive and process information and the efficiency of information dissemination.

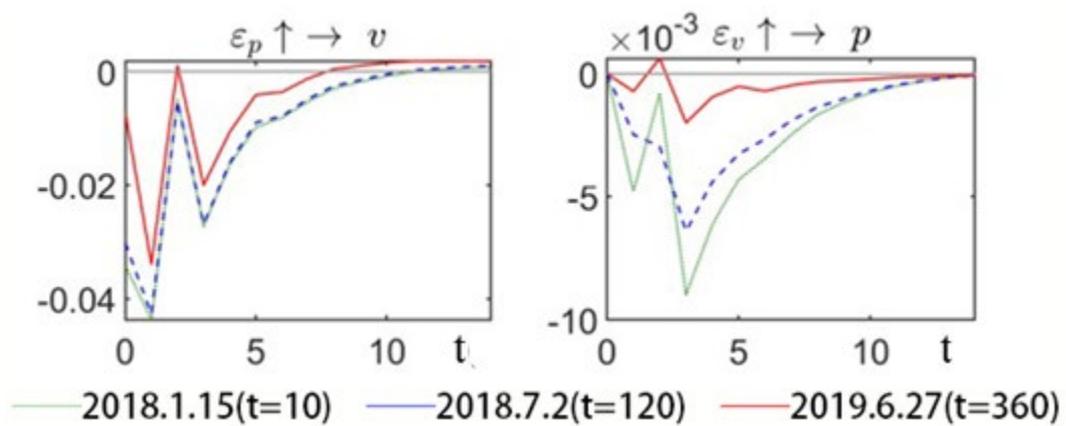

Figure 7 Impulse response between investor sentiment and

stock market volatility at different points in time

# 5 Conclusions and recommendations

This paper makes a study on the impact of investor sentiment on the stock market and proposes the use of a deep learning approach to study a large number of market sentiments. Based on the data of EastMoney and SZCZ, investor sentiment is constructed using the BERT model. Market liquidity and market volatility indicators are also introduced. The experimental results show a high correlation, indicating that investor sentiment does have a large impact on the stock market. The remaining deviations in the results may be caused by the high noise of the stock bar data in the empirical analysis and minor deviations in the analytical model. However, the hypothesis is met on the whole and the experimental results illustrate the strong robustness of the model.

The policy recommendations based on this study are: **a)** Investor sentiment in the Chinese stock market has a large impact on the stock market, especially individual investor sentiment which is easily influenced by sentiment on social media platforms, which makes a requirement for further improvement of the national market economy system. With the improvement of the stock market and the increase of institutional investors, the sensitivity of stock prices to investor sentiment will be reduced and market risk will be reduced. **b)** While making the development of institutional investors an important development strategy for China's stock market, the issue of how to improve the quality of institutional investors and enhance the rationalisation of their investment behaviour is an equally important issue. If this problem is solved well, the quality of individual investors can be improved, and thus the efficiency of the market can be improved.

There are some imperfections in this paper that need to be further studied by scholars. For example, future work could try to study the impact of uncertainties such as national policies, markets, and unexpected events. More explanatory variables can also be added to the model to reduce noise, and long-period samples and smaller samples can be used to further experiment with the model and parameters to improve the performance of the model.

[34] Ren YP,Li W. Economic policy uncertainty, investor sentiment and stock pricesynchronization - time-varying parameters based on TVP-VAR model[J]. Journal of Shanghai University (Natural Science Edition),2020,26(05):769-781.